 \documentclass[final,authoryear,5p,times,twocolumn,10pt]{elsarticle}

 \usepackage{graphicx}

\usepackage{amssymb}
 \usepackage{amsthm}

\usepackage{graphics}
\usepackage{amsmath}
\usepackage[QX,T1]{fontenc}
\usepackage[english]{babel}
\usepackage[latin1,utf8]{inputenc}
\usepackage{graphicx}
\usepackage{eepic}
\usepackage{latexsym}
\usepackage{float}
\usepackage{textcomp} 

\newcommand{\un}[1]{\mbox{ \rmfamily #1}}
\newcommand{\unp}[1]{\mbox{\rmfamily #1}}

\newcommand{\unrho}[0]{\mbox{ \rmfamily kg} / \mbox{\rmfamily m} ^3}

\newcommand{\url}[1]{\ttfamily #1\normalfont}
\newcommand{\fig}[1]{Fig.~\ref{#1}}
\newcommand{\tab}[1]{Table~\ref{#1}}

\newcommand{\sect}[1]{Sect.~\ref{#1}}

\newcommand{\undeg}{\mbox{\textdegree}}
\newcommand{\unmin}{'}
\newcommand{\unsec}{''}

\journal{Icarus}

\begin{document}

\begin{frontmatter}



\title{The observing campaign on the deep-space debris WT1190F as a test case for short-warning NEO impacts}


\author[NEOCC,OAR]{Marco Micheli\corref{Phone: +39 06 941 80 365}}
\ead{marco.micheli@esa.int}
\author[OABO]{Alberto Buzzoni}
\author[NEOCC,ESTEC,TUM]{Detlef Koschny}
\author[NEOCC,Oldenburg]{Gerhard Drolshagen}
\author[ASI,NEOCC,Deimos-RO]{Ettore Perozzi}
\author[ESO]{Olivier Hainaut}
\author[SDO]{Stijn Lemmens}
\author[OABO,OAR]{Giuseppe Altavilla}
\author[OABO]{Italo Foppiani}
\author[Deimos-ES]{Jaime Nomen}
\author[Deimos-ES]{Noelia S\'anchez-Ortiz}
\author[Lumezzane]{Wladimiro Marinello}
\author[Lumezzane]{Gianpaolo Pizzetti}
\author[Lumezzane]{Andrea Soffiantini}
\author[Purdue]{Siwei Fan}
\author[Purdue]{Carolin Frueh}

\address[NEOCC]{ESA SSA-NEO Coordination Centre,
Largo Galileo Galilei, 1, 00044 Frascati (RM),
Italy}

\address[OAR]{INAF - Osservatorio Astronomico di Roma,
Via Frascati, 33, 00040 Monte Porzio Catone (RM),
Italy}

\address[OABO]{INAF - Osservatorio Astronomico di Bologna,
Via Gobetti, 93/3, 40129 Bologna (BO),
Italy}

\address[ESTEC]{ESTEC, European Space Agency,
Keplerlaan 1, 2201 AZ Noordwijk,
The Netherlands}

\address[TUM]{Technical University of Munich,
Boltzmannstra\ss e 15, 85748 Garching bei München,
Germany}

\address[Oldenburg]{Space Environment Studies - Faculty VI, Carl von Ossietzky University of Oldenburg,
26111 Oldenburg,
Germany}

\address[Deimos-RO]{Deimos Space Romania,
Strada Buze\c{s}ti 75-77, Bucure\c{s}ti 011013,
Romania}

\address[ASI]{Agenzia Spaziale Italiana,
Via del Politecnico, 1, 00133 Roma (RM),
Italy}


\address[ESO]{European Southern Observatory,
Karl-Schwarzschild-Stra\ss e 2, 85748 Garching bei München,
Germany}

\address[SDO]{ESA Space Debris Office,
Robert-Bosch-Stra\ss e 5, 64293 Darmstadt,
Germany}

\address[Deimos-ES]{Deimos Space S.L.U.,
Ronda de Pte., 19, 28760 Tres Cantos, Madrid, 
Spain}

\address[Lumezzane]{Osservatorio Astronomico ``Serafino Zani'',
Colle San Bernardo, 25066 Lumezzane Pieve (BS),
Italy}

\address[Purdue]{School of Aeronautics and Astronautics, Purdue University,
701 W Stadium Ave, West Lafayette, IN 47907,
USA}

\begin{abstract}
On 2015 November 13, the small artificial object designated WT1190F entered the Earth atmosphere above the Indian Ocean offshore Sri Lanka after being discovered as a possible new asteroid only a few weeks earlier. 
At ESA's SSA-NEO Coordination Centre we took advantage of this opportunity to organize a ground-based 
observational campaign, using WT1190F as a test case for a possible similar future event involving 
a natural asteroidal body.
\end{abstract}

\begin{keyword}
Near-Earth objects \sep Asteroids, dynamics
%

\end{keyword}

\end{frontmatter}



\section{Introduction}

The object known with the temporary observer-assigned designation of WT1190F was discovered 
by the Catalina Sky Survey on 2015 October 03 \citep{matheny15}, and quickly reported for 
confirmation as a 
possible new Near-Earth Object (NEO) to the Minor Planet Center. It was quickly realized to be in a geocentric orbit by the JPL Scout system \citep{Farnocchia15a,Farnocchia16}, and after a few days of follow-up it 
was identified to also be on a collision trajectory with Earth, with a predicted impact on the morning of 2015 November 13, offshore the coastal region of Southern Sri Lanka \citep{jenniskens16}.

Even with a short arc of observations it quickly became clear that the object, just a few 
meters in diameter, was subject to significant non-gravitational perturbations due to the 
effects of radiation pressure \citep{gray15}. The magnitude of such effect implied an extremely low mean 
density, of the order of $100\unrho$, sufficient to exclude a truly asteroidal nature and 
suggesting a man-made origin, such as a hollow shell remnant of some unidentified spacecraft.

Although it was then clear that the object was not natural, and therefore not a real near-Earth 
asteroid, the team of ESA's SSA-NEO Coordination Centre decided to use this opportunity to 
organize an observational campaign in the shortest possible time, as a simulation and training 
for what kind of data would be possible to obtain in case of a future similar event involving 
a natural impactor.

The following is a report of this effort, focusing on the type of observations that were 
obtained, and which telescopes we were able to access during the available time. Our goal 
is to show that, through an effort of coordination of worldwide astronomical resources, it is 
possible to make the best possible use of available assets to obtain a complete set of 
observations that would be sufficient to fully characterize the event. For the actual scientific 
analysis of such data we redirect the reader to the exhaustive work of \citet{buzzoni17a,buzzoni17b}.


\section{Observations}

In the following sections we will quickly summarize the types of observations we were able 
to obtain, either directly or through collaborations, in the relatively short timespan between the 
discovery of WT1190F and its impact with Earth (much shorter than the typical cycle of telescope proposals).

It is important to note that, due to the unpredictable nature of these discoveries, none of 
these observations could have been planned in advance and requested to professional telescopes 
via the standard form of a regular telescope proposal.
Most of them were obtained through collaborations with small and medium size observatories, 
which allow for the necessary flexibility and rapidity of response that were essential in this 
case.
We are however pleased to report that even the more traditional channel of a Director's 
Discretionary Time (DDT) request for urgent observations at a large aperture telescope, like
the ESO VLT at Cerro Paranal (Chile) 
was successful, showing that there is a possibility to obtain access to those 
highly competitive instruments (for small amounts of time) if the target is sufficiently 
urgent and unique.


\subsection{Astrometric follow-up}\label{Astrometry}

The very first type of observation that is important to obtain shortly after the discovery 
of a new moving object is astrometric follow-up. The need for astrometric observations is 
even more urgent for an object in a collision path, because an accurate determination 
of its trajectory is essential to properly predict the impact point and time. More indirectly, 
for a small object like WT1190F, good astrometric coverage is essential to quickly ascertain 
the effects of solar radiation pressure on its dynamics, and determine the ratio between its 
cross sectional area and mass (the so called ``Area to Mass Ratio'', or AMR), which can in 
turn be converted into an estimate of its density, providing hints on its nature.

\begin{figure}
\centering
    \includegraphics[width=0.85\columnwidth]{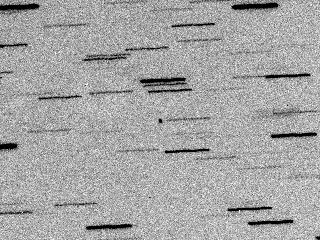}
  \caption[Trailed image]
  {An illustrative astrometric image, taken along the night of 2015 November 12,
  just a few hours before the WT1190F atmosphere entry, with the $1.52\un{m}$ ``Cassini'' telescope of the
  Loiano Observatory (Italy). The telescope was tracking non-sidereally at the apparent angular 
  motion of the target (which appears as a point source near the center of the field), 
  therefore all astrometric reference stars 
  in the frame are severely trailed (approximately $1\unmin$ in this example). A correct 
  astrometric reduction of a frame like this requires to fit every reference star with a trailed 
  model, to properly determine their centroid.}
  \label{Trails}
\end{figure}

For this reason, in the few days after discovery we contacted collaborators in our network of 
observers to request images of the object, providing them with the most up-to-date station-specific 
ephemerides we generated internally with the software {\sc Find\_Orb} by Bill Gray\footnote{The {\sc Find\_Orb} programme is
publicly available at the URL:\\ {\url{http://www.projectpluto.com/find\_orb.htm}}.}
Over the few weeks between discovery and impact we were able to obtain images from various 
observatories, including those of Asiago and Loiano, both managed by the Italian 
Institute for Astrophysics (INAF), and ESA's own Optical Ground Station in Tenerife (Spain).
We extracted high precision astrometry from each image set, using tools that can properly manage 
images where either the field stars, or the moving object (or both) are severely trailed 
(see \fig{Trails} for an example of such images). For each telescope, we also carefully 
investigated with the observer the presence of possible time biases, which may have dramatic 
effects on the trajectory determination for objects moving at high high plane-of-sky rates.

The astrometry resulting from these measurements was all submitted to the Minor Planet Center 
(MPC), and quickly published in their DASO (Distant Artificial Satellites Observation) 
circulars. However, for each position we also maintained record of our own positional and 
timing uncertainties, that cannot yet be included in the MPC astrometric format, but are 
nevertheless extremely valuable during the orbit determination process.

This dataset, together with the pre-impact data presented in \sect{Impact} formed the basis of 
our trajectory and impact analysis, which is fully presented in \citet{buzzoni17a}. This high 
precision orbit was also essential to guarantee that later observations, such as the slit 
spectroscopy discussed below in \sect{Spectroscopy}, could be properly carried out.


\subsection{Pre-impact follow-up}\label{Impact}

In addition to the astrometric coverage obtained during the weeks between discovery and impact, 
in a case like WT1190F it is important to obtain observations during the last hours before the 
impact event. These observations are extremely useful for trajectory determination, because the 
reduced distance allows for a much better spatial accuracy for a given angular resolution.
However, during the final phases of the approach, an object like WT1190F can become extremely 
fast, reaching an angular speed in excess of $1000\unsec/\unp{min}$ where the accuracy in the 
timing signals used to timetag the images becomes the leading source of error in the astrometric 
measurement. Furthermore, when the object is only a few 
thousands of kilometers away, even the accuracy in the geographical coordinates of the telescope 
becomes essential.

During the night between 2015 November 12 and 13 UT we tried to obtain the most complete 
possible coverage of the object. Our group directly used the $1.52\un{m}$ reflector in Loiano 
(MPC code 598) and the $0.40\un{m}$ reflector in Lumezzane (Italy, MPC code 130) to collect continuous 
observations and perform astrometric reductions in near real time. Additional astrometry was 
also provided by the {\sc Deimos} team from its Mt.\ Niefla DeSS Observatory in Spain 
(MPC code Z66), through a couple of $0.40\un{m}$ and $0.28\un{m}$ reflectors, which further 
helped refine the trajectory. 
All these observations continued until less than an hour before impact, when the object set and 
twilight began to interfere with the observations.

\begin{table*}
\centering
\begin{tabular}{lcrrcccrl}
\hline
Telescope &    Date range     & \# Obs. & Range distance &  \multicolumn{2}{c}{Ballistic impact point} & Impact time & \multicolumn{2}{c}{Error budget} \\
 (MPC ID) &    days of 2015 Nov &         & \multicolumn{1}{c}{(km)} & N Lat. (\undeg) & E Lon. (\undeg)  &  UTC       & Position & Timing \\
          &                   &         &                          &                     &                     &  hh:mm:ss  & (m)      & (s)  \\       
\hline
\noalign{\smallskip}
309       & 06.25 $\to$ 06.28 &   3 & \multicolumn{1}{c}{$574\,000$}    &  ...      &  ...        &  ... &   ... &  ... \\
598       & 08.06 $\to$ 13.17 &  12 & $518\,000 \to 20\,000$ & +5.63172 & +81.53236 & 06:18:49.79 & 1500 &  1.8  \\	
130       & 12.97 $\to$ 13.23 &  20 & $ 95\,000 \to 15\,000$ & +5.63249 & +81.52863 & 06:18:49.71 & 1500 &  1.7  \\	
Z66       & 13.17 $\to$ 13.24 & 348 & $ 38\,000 \to 13\,000$ & +5.63099 & +81.53689 & 06:18:49.87 &  500 &  1.4  \\	
309+598   & 06.25 $\to$ 13.17 &  15 & $574\,000 \to 20\,000$ & +5.63132 & +81.53271 & 06:18:50.02 & 1000 &  1.0  \\
598+130   & 08.06 $\to$ 13.23 &  32 & $518\,000 \to 15\,000$ & +5.63409 & +81.52367 & 06:18:49.53 &  200 &  0.4  \\	
598+Z66   & 08.06 $\to$ 13.24 & 360 & $518\,000 \to 13\,000$ & +5.63368 & +81.52470 & 06:18:49.77 &  150 &  0.4  \\
\hline
\noalign{\smallskip}
All above & 06.25 $\to$ 13.24 & 363 & $574\,000 \to 13\,000$ & +5.63252 & +81.52816 & 06:18:49.83 &   25 &  0.05 \\   
JPL       &                   &     &                        & +5.62756 & +81.49063 & 06:18:49.21 &   90 &  0.03 \\
\hline
\end{tabular}
\caption{Estimated impact location at ground level assuming a purely ballistic entry (no atmospheric drag), computed from different subsets of astrometric data. Dates are in decimal UT days of 2015 November, 
times are UT of 2015 November 13. The table uses IAU codes for each station: 
130=Lumezzane, 309=Paranal (VLT), 598=Loiano, Z66=DeSS. The nominal prediction from the
JPL final trajectory reconstruction \citep{farnocchia15} is reported in the last entry, for comparison purposes. Error bars quoted in the table are semiaxes of the uncertainty ellipses at the $1\sigma$ level.}
\label{DASOastrometry}
\end{table*}

In \tab{DASOastrometry} we present a summary of the information that can be extracted from a selection 
of the astrometric datasets we obtained, analyzed with the {\sc Find\_Orb} software.
When combined together, our ground-based astrometric observations allow for an extremely accurate 
determination of the impact circumstances, at least to the level of the highest layers of 
atmosphere. The formal uncertainty in time along track is better than $0.1\un{s}$, while the impact point 
at a level of $100\un{km}$ above the surface can be determined with an accuracy of about $100\un{m}$.
Clearly, this formal uncertainty does not directly correspond to the prediction accuracy for the actual 
impact point with the Earth's surface, because of the catastrophic effects happening on the object 
during the atmospheric entry phase. Nevertheless, they provide an indication of how well ground-based 
observations can constrain the trajectory of an incoming object, to a level where other 
factors, such as atmospheric dynamics, fragmentation height and physical characteristics of the 
body become by far the leading source of unmodeled uncertainty in the re-entry 
dynamics\footnote{It is important to point out that, due to the moderately steep incident angle
of $\sim 20\undeg$, WT1190F behaved far more like an 
asteroid than a man-made spacecraft in its atmosphere re-entry. The object actually approached 
Earth at a 10.6 km s$^{-1}$ velocity, thus crossing our atmosphere in some $\sim 10\un{s}$ 
\citep{jenniskens16}.
As consequence of the limited aerodynamic drag interval the true impact point should have expected 
to be earlier along the track than predicted by a purely ballistic model, but not too significantly so.}.

It is interesting to note that a single observatory collecting data until shortly before impact 
(DeSS) can provide a slightly better accuracy to a longer temporal coverage but terminated earlier 
(Loiano). On the other hand, an obvious strategic advantage can be envisaged for the latter case, 
as earlier (deeper) observations with intermediate-class telescopes would assure a greatly 
anticipated alert in case of forthcoming impact threat.
Combining two datasets, like Loiano+DeSS, provides of course the ideal solution, as 
we can fully take advantage of the parallax effect between the two stations. Using multi-station 
astrometry and a longer arc results in an even better accuracy in both spatial and temporal 
direction.

\subsection{Ligthcurve observations}\label{Lightcurves}

Another important component of any effort to characterize hazardous objects approaching
Earth relies on ground-based lightcurve photometry, useful to determine the rotational period 
of the body, which in turn may provide another clue on its nature and origin.
In the case of WT1190F, for instance, the increased luminosity of the object while
approaching Earth in its final trajectory allowed accurate fast photometry to be carried out from
our ground-based telescope network, thus revealing that the body was
in fact an extremely fast rotator, with a period $P = 1.455_{\pm 0.001}$ s and a complex geometry.
A thorough discussion of these results, and their relevance to constrain
the dynamical properties of this puzzling target are presented in \citet{buzzoni17a,buzzoni17b}. 
It is just worth stressing here that a fast-spinning body could more likely survive the 
atmosphere entry event allowing heat to be more easily dissipated, such as to avoid (or limit) 
any body disruption/fragmentation.
These arguments actually led \citet{jenniskens16} to conclude that about two thirds of 
WT1190F's entire mass may have safely arrived at the ground, impacting the sea still at 
supersonic speed. 

For the goals of the present discussion, however it is important to point out that a search 
for photometric periodicities in this kind of targets may be extremely time 
consuming, and needs to be planned in advance to have access to appropriate facilities for long 
enough timespans. Natural objects can sometimes have rotational periods up to many hours (or 
occasionally days), while artificial objects may often rotate with periods of the order of seconds
\citep[e.g.][]{frueh10,cowardin12,hall14}.

To further complicate the analysis, it is not unusual for both natural and artificial objects in 
this size range to be in complex rotational states, as it was directly observed by 
\citet{scheirich10} for 2008 TC3, the first natural Earth impactor discovered before impact.

\subsection{Spectroscopy and colors}\label{Spectroscopy}

Spectroscopy is often the most direct way to remotely gather information on the physical properties, 
composition and nature of an object \citep[e.g.][]{vananti17}. It is therefore extremely important 
to obtain some kind of spectroscopic information (or its proxy in term of multicolour photometry) on an 
impacting object as early as possible, to ascertain its nature.

A reflectance spectrum is first of all useful to prove the natural or artificial nature of the 
body. If an object turns out to be artificial, a spectrum can provide information on the material, 
alloy or painting used on the body. Even more importantly, if the body turns out to be natural, 
a spectrum would allow for a taxonomical classification of the NEO, which would in turn 
provide information on its composition, density, size and properties of the material, essential 
to estimate the possible ground damage in case the collision turns out to be threatening for 
population or assets on the ground.

From a more scientific perspective, even in case of a smaller object, the possibility to compare 
a taxonomical classification obtained with the object still in space with any meteorite collected 
on the ground afterward is very valuable. This for example became extremely important when 
2008 TC3, observed to be an F-type in space, turned out to have predominantly ureilitic 
composition when meteorites were collected on the ground, thus providing evidence for a linkage 
between F-type asteroids and ureilites \citep{jenniskens09}.

Obtaining a good spectrum of an object like WT1190F days before impact was not trivial, since 
the object was typically keeping a magnitude $V\sim21$, thus requiring large aperture 
instruments. We therefore decided to submit a DDT proposal at the ESO telescopes for
$30\un{min}$ of FORS2 time on VLT to obtain visible spectroscopy of the object around its
last apogee transit in early November of 2015.
Our preliminary follow-up, including our own astrometric observations, was necessary to 
achieve a convenient accuracy in the ephemeris prediction ($< 20\unsec$) for   
the target to be safely tracked within the $1\unsec$ slit of the spectrograph.
The non-sidereal tracking capabilities of VLT were also necessary to maintain the object properly 
aligned with the slit for the whole exposure time at speeds as high as $5\unsec/\unp{min}$ that 
were typical of the object a few weeks before impact.

\begin{figure}
\centering
    \includegraphics[width=\columnwidth]{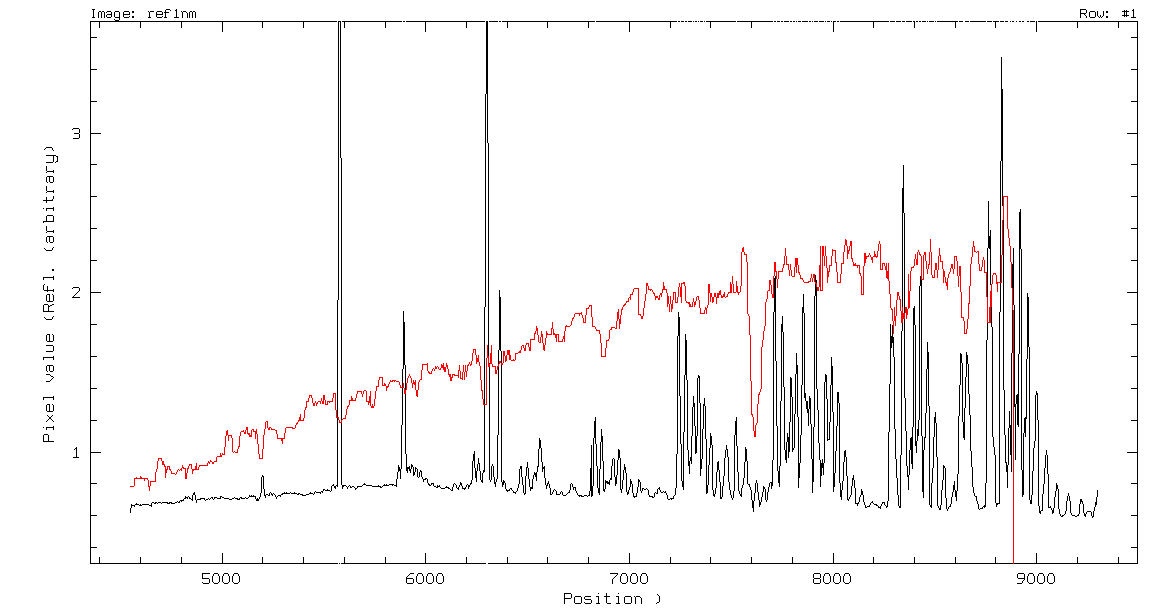}
  \caption[VLT spectrum]
{The sky-subtracted spectrum of WT1190F obtained on 2016 November 06 with the FORS2 
camera of the ESO VLT telescope (in red) with the superposed sky background (in black). 
The absolute calibration allowed to assess WT1190F's inherent colors, which resulted only 
slightly redder than the Sun and closely matching those of a typical red giant star of 
spectral type K3.
}
  \label{Spectrum}
\end{figure}

The observations were successfully carried out along the night of 2015 November 6
(see \fig{Spectrum}), that is 
within a week from the proposal submission, and allowed us to get a quick assessment of the 
spectral properties of WT1190F many days before it arrived to Earth. 
According to its apparent spectral energy distribution, the object was only slightly 
''redder'' than the Sun, and more closely resembled a star of spectral type K3. These conclusions
were actually corroborated the following night of 2015 November 7, where multicolour
observations from Loiano in the Johnson-Cousins photometric system confirmed a color 
$B-V = 1.00_{\pm 0.23}$ and $V-R_c = 0.52_{\pm 0.24}$ \citep{buzzoni17a}.

Additional spectroscopic information, fully supporting these results, were also obtained a few days 
later with the BFOSC camera of the ''Cassini'' telescope from Loiano \citep{buzzoni17a}, 
when the object was already significantly brighter and easier to observe with 
smaller apertures. It is however important to point out that the VLT observations, obtained 
and reduced more than a week before impact, would have been essential in case the object would 
have presented an actual impact threat, because they would have allowed for enough time to 
organize a mitigation effort on the ground.

\subsection{Re-entry observations and campaigns}

In addition to the observational results presented above, our team of the SSA-NEO Program was 
also able to provide funding for two European researchers (Stefan L\"ohle and Fabian Zander of 
the University of Stuttgart) to participate in the international airborne observation campaign 
sponsored by the International Astronomical Center in Abu Dhabi and the United Arab Emirates Space
Agency, which flew over the impact location in Sri Lanka and directly observed the re-entry. 
A full report of this mission and the relevant results has been presented by \citet{jenniskens16}.

Although not directly relevant to this analysis, the possibility of organizing such 
collaboration is also a clear example of how quickly an international effort like this can be 
organized and successfully carried out.

\section{Discussion}

The observations obtained in this campaign prove that, to achieve complete observational 
coverage of an incoming object on short notice, it is necessary to develop a network of 
instruments of different classes, each of which is capable of providing a given type observation 
covering a specific niche in the overall observational strategy.

In particular, it is often necessary to have access to:
\begin{itemize}
\item At least one large telescope (e.g. VLT), which allows for cutting-edge observations 
(such as spectroscopy at very faint magnitude levels) early in the approach phase. At the same 
time, being an expensive resource, a telescope of this class can only be obtained on short 
notice for a limited amount of time.
\item One or two mid-class telescopes (e.g. Loiano), which can be used with much more 
flexibility, for multiple nights, but may not be sensitive enough at large distances.
\item A network of small telescopes (e.g. Lumezzane), which can be triggered even on very 
short notice and made available for extended period of time, and are ideal to cover the 
last phases of the impact trajectory from multiple locations.
\end{itemize}

From a planning perspective, the WT1190F campaign also showed that a complete coverage can 
only be achieved with a 2-phase strategy, including:

\begin{itemize}
\item A ``long arc'' observational coverage at large geocentric distances (i.e. trans-lunar), 
carried out with middle to large class professional telescopes, and mostly dedicated to the
collection of immediate follow-up data and physical information
\item A ``short leg'' refining phase to be executed in the days or hours before impact, where 
astrometric quality and number of observing stations are the key factor to ensure the best 
knowledge of the impact trajectory, while aperture becomes less important thanks to the 
increased brightness of the target.
\end{itemize}

\section{Conclusions}

The case of WT1190F, the first Earth impactor discovered with more than a day of advance 
notice, provided an ideal real-life test case for how to quickly organize an observational 
campaign with multiple instruments and observational techniques.

Our results presented above shows that a few weeks are sufficient to provide a full 
observational coverage with a wide array of observational techniques.
For some types of measurements, such as astrometry, both quick reaction time and extended 
coverage are important factors, but even modest-sized telescopes can often provide very 
valuable data, assuming their setup is well controlled and understood (in terms of tracking 
and timing capabilities).
Other observations, and spectroscopy in particular, require a much greater sensitivity, 
which can be provided only with large aperture instruments. In these cases, the most suitable 
channel is to make use of DDT opportunities made available by professional observatories, 
which can be successful if submitted in a timely fashion.

It is nevertheless important to point out that the case of WT1190F may have been unusual 
compared to a more typical small Earth impactor, due to the comparatively long interval 
between discovery and impact. Since WT1190F was actually an Earth-orbiting body, it was 
possible to discover it before its incoming plunge trajectory, and this extra time was 
very useful to get a more complete analysis of the body.

In the case of a true heliocentric impacting asteroid of the same size, it is likely that 
the warning time may be lower, of the order of a few days or less. 
It is likely that sufficient astrometric coverage will still be possible in this case, and adequate coverage of the impacting trajectory could also be achieved, providing that the impact is known with at least a few hours of advanced warning. However, some other observations, such as the full spectral coverage we obtained with VLT, require an advance notice of about a week, which may not have been possible for an object of this size which was coming towards Earth in an hyperbolic trajectory.

An advance notice of about a week, and a magnitude brighter than $V\sim21$, are typically needed to perform most of the observations we obtained on WT1190F. For an average hyperbolic impactor discovered early enough by a survey, and recognized as such, these thresholds could be achievable for objects of $H\sim27$ or larger. This size range nicely corresponds to the size limit above which mitigation or evacuation efforts may become useful (e.g. the size of the Chelyabinsk impactor).
We can therefore expect that the experience, methods and contact points developed for this campaign will become an essential 
resource in case one of these campaigns will need to be organized in the future, especially if timing constraints were more stringent.

\section*{Acknowledgments}

We would like to thank all the observers and collaborators who contributed the observations 
that were presented in this analysis: D.~Abreu, S.~Benetti, I.~Bertini, 
B.~Bolin, I.~Bruni, M.~Busch, A.~Coffano, R.~Gualandi, A.~Kn\"ofel, M.~Lazzarin, 
M.~Masserdotti, P.~Ochner, P.~Ruiz, E.~Schwab and R.~Wainscoat.


\bibliographystyle{model2-names}

\end{document}